\documentclass[conference]{IEEEtran}
\usepackage{amsmath,amssymb,amsfonts}
\usepackage{graphicx}
\usepackage{textcomp}
\usepackage{xcolor}
\usepackage[colorlinks=true,allcolors=blue]{hyperref}
\DeclareMathOperator*{\argmax}{arg\,max}
\newcommand{\C}{\mathbb{C}}

\newcommand{\Gr}{\mathcal{G}}
\newcommand{\bx}{\mathbf{x}}
\newcommand{\by}{\mathbf{y}}
\newcommand{\bh}{\mathbf{h}}

\newcommand{\bW}{\mathbf{W}}
\newcommand{\bY}{\mathbf{Y}}

\begin{document}

\title{A Low-PAPR, Synchronization-Robust\\ Non-Coherent Grassmannian Modulation\\ for Optical Communications}

\author{\IEEEauthorblockN{Eylon E. Krause$^{1,2}$}
\IEEEauthorblockA{eylonkr@colman.ac.il \\ Weizmann Institute of Science$^1$, Rehovot, Israel (7610001)  \\ College of Management Academic Studies$^2$, Rishon LeZion, Israel (7579806) }}

\maketitle
\pagestyle{plain}
\thispagestyle{plain}

\begin{abstract}
Non-coherent Grassmannian (unitary space--time) signaling detects on the
received \emph{subspace}, which is invariant to a branch-side unitary on the
receive dimension (a polarization or, under space-division multiplexing, a
mode-coupling rotation) and to a phase that is constant over the coherence
block (the common laser-phase contribution). It therefore needs no
carrier-phase or polarization recovery \emph{within the block} and is robust
to phase noise as long as the phase drift across the block is small, while a
multi-branch (polarization- or space-division-multiplexed) front end harvests
diversity without channel estimation or pilots. However, the
Grassmannian-constellation literature almost universally assumes two conditions
that fail on optical links: a \emph{distortion-free, linear} channel and
transmitter, and \emph{already-acquired symbol timing}. This paper closes both
gaps while reusing off-the-shelf Grassmannian packings. First, we impose a
constant-modulus (low peak-to-average-power-ratio, PAPR) constraint on the
constellation and quantify the resulting trade-off between PAPR and chordal
distance: a constant-modulus design lowers the $0.1\%$ PAPR from $6.1$~dB
(unconstrained) to $3.6$~dB---$1.6$~dB below $16$-QAM ($5.2$~dB)---which eases
the optical modulator's linear range and is expected to reduce the fiber
Kerr-nonlinearity penalty, at a $\sim\!1.8$~dB cost in high-SNR coding gain.
Second, we derive a \emph{phase-blind subspace timing-error detector} (TED)
that exploits the invariance of the GLRT projection energy to the unknown
carrier phase, together with a feedforward acquisition metric---supplying clock
recovery without prior carrier or polarization recovery. The TED yields a clean
S-curve with a stable lock point for roll-off factors down to $\beta=0.1$.
Under block fading the proposed estimator attains genie-timing
symbol-error-rate within a fraction of a dB and recovers full diversity,
whereas an uncorrected $0.35$-symbol timing offset floors the error rate near
$0.4$. All link-level results use a symbol-rate block-fading/AWGN abstraction,
with full fiber, modulator, and laser-phase-noise modeling left to future work.
The scheme combines low PAPR with the diversity and (within-block)
phase-recovery-free operation of non-coherent reception.
\end{abstract}


\section{Introduction}
Coherent optical transceivers achieve high spectral efficiency, but their
digital signal processing (DSP) chain---chromatic-dispersion compensation,
polarization demultiplexing, carrier-frequency and carrier-phase recovery, and
clock recovery---rests on tracking a time-varying channel \cite{ip,savory}.
Two of those tracking loops are persistently stressed: carrier-phase recovery
must follow laser phase noise, which tightens sharply for higher-order formats
and low-cost (wide-linewidth) lasers \cite{pfau}, and polarization recovery
must follow fast state-of-polarization transients, an effect that grows in
mode-coupled space-division-multiplexed (SDM) fibers \cite{richardson}. A
complementary physical-layer constraint is that the transmitter and fiber are
nonlinear: the Mach--Zehnder modulator has a limited linear range and the
Kerr effect penalizes high peak powers, so low peak-to-average-power-ratio
(PAPR) waveforms are preferred.

Non-coherent \emph{Grassmannian} signaling---also called unitary space--time
modulation (USTM)---offers an alternative that sidesteps the tracking loops. A
symbol is a point on the Grassmann manifold $\Gr(p,T,\C)$ of $p$-dimensional
subspaces of $\C^{T}$, and detection depends only on the received
\emph{subspace}, which is invariant to a branch-side unitary on the receive
dimension (a polarization or, in SDM, mode-coupling rotation) and to a phase
common to all $T$ slots of the block (the laser-phase contribution that is
constant over the block) \cite{hochwald,zheng}. This buys two properties
optical links want: (i) diversity equal to the number of receive branches
(polarization or spatial), harvested without channel estimation or pilots, and
(ii) immunity to a constant per-block carrier phase, hence no carrier or
polarization recovery within the coherence block. A \emph{time-varying} phase
across the block---a carrier-frequency-offset ramp or within-block (Wiener)
phase-noise drift---is not absorbed by the subspace and leaves a residual floor
that shrinks as the per-block phase drift, and hence $T$, decreases. A mature
body of work supplies near-optimal packings and structured, low-complexity
families, e.g., Conway--Hardin--Sloane packings \cite{chs}, Cube-Split
\cite{cubesplit}, Grass-Lattice \cite{grasslattice}, and exponential-map codes
\cite{kammoun}.

The gap this paper addresses is that essentially all of this work assumes a
\emph{linear} channel and transmitter and \emph{block-synchronous} reception:
the receiver is handed the symbol timing and only the channel/phase is unknown.
Both assumptions fail on optical links. Our contributions are:
\begin{enumerate}
\item \textbf{Low-PAPR Grassmannian for the optical link.} Constant-modulus
(constant-amplitude) Grassmannian symbols are themselves known \cite{covert};
our contribution is to apply the constraint in the optical setting and to
quantify the resulting PAPR/chordal-distance trade-off, so the pulse-shaped
envelope eases the modulator's linear range and is expected to reduce the
fiber Kerr-nonlinearity penalty (Sec.~\ref{sec:papr}).
\item \textbf{Subspace-matched timing recovery.} We adapt non-data-aided
square-law timing recovery to the Grassmannian subspace-detection (GLRT
projection) statistic, giving a carrier-phase-independent TED matched to the
subspace rather than to a symbol decision, analyze its S-curve, and give a
feedforward acquisition metric (Sec.~\ref{sec:timing}). Phase-blind clock
recovery is itself classical \cite{gardner,oerder}; what the
Grassmannian-constellation literature lacks---and what we supply---is a timing
detector built on the subspace statistic.
\item \textbf{General performance evaluation.} We benchmark error rate with
genie, estimated, and uncorrected timing under block fading, and contrast PAPR
against $16$-QAM (Sec.~\ref{sec:results}).
\end{enumerate}

\section{System Model}\label{sec:model}
We consider an optical link observed over $N$ diversity branches---e.g., the
two polarizations, or $N$ spatial/modal channels in an SDM fiber---modeled as a
block-fading channel.
Over one coherence block of $T$ channel uses the transmitter sends a unit-norm
vector $\bx\in\C^{T}$, $\|\bx\|=1$, drawn from a constellation
$\mathcal{X}=\{\bx_1,\dots,\bx_M\}$. The received matrix
$\bY\in\C^{T\times N}$ is
\begin{equation}
\bY=\bx\,\bh^{\mathsf H}+\bW,
\label{eq:channel}
\end{equation}
where $\bh\in\C^{N}$ collects the per-branch complex gains
($h_n\!\sim\!\mathcal{CN}(0,1)$, unknown), carrying the per-branch amplitude
fading together with the phase that is \emph{constant over the block} (the
common laser-phase term and any branch-side polarization/mode rotation), and
$\bW$ has i.i.d.\ $\mathcal{CN}(0,N_0)$ entries. Two idealizations are worth
naming: the i.i.d.\ $\mathcal{CN}(0,1)$ model treats the branches as
independently Rayleigh-faded, whereas real polarization/SDM branches are
coupled by a near-unitary (power-preserving) rotation, so the order-$N$
diversity claimed below is an i.i.d.-Rayleigh upper bound; and a residual
carrier-frequency offset or within-block phase drift, being time-varying across
the $T$ slots, is \emph{not} captured by the single per-block $\bh$. Because
$\bh$ carries an unknown phase and magnitude, the maximum-likelihood /
generalized-likelihood-ratio detector for equal-energy codewords reduces to a
\emph{subspace projection}:
\begin{equation}
\hat{m}=\argmax_{m}\ \bx_m^{\mathsf H}\!\left(\bY\bY^{\mathsf H}\right)\bx_m
       =\argmax_{m}\ \sum_{n=1}^{N}\bigl|\bx_m^{\mathsf H}\by_n\bigr|^2,
\label{eq:glrt}
\end{equation}
with $\by_n$ the $n$-th column of $\bY$. Symbols are compared by the
\emph{chordal distance}
$d_c(\bx_i,\bx_j)=\sqrt{1-|\bx_i^{\mathsf H}\bx_j|^2}$, and at high SNR the
pairwise error probability decays as
$\big(d_c^2\,\rho\big)^{-N}$: the diversity order equals $N$ and, by a
union-bound argument, the coding gain is governed by the minimum chordal
distance $d_{c,\min}=\min_{i\neq j}d_c(\bx_i,\bx_j)$ \cite{hochwald,zheng}. The
rate is $R=\tfrac{1}{T}\log_2 M$ bits/channel use.

\begin{figure}[t]
\centering
\includegraphics[width=\columnwidth]{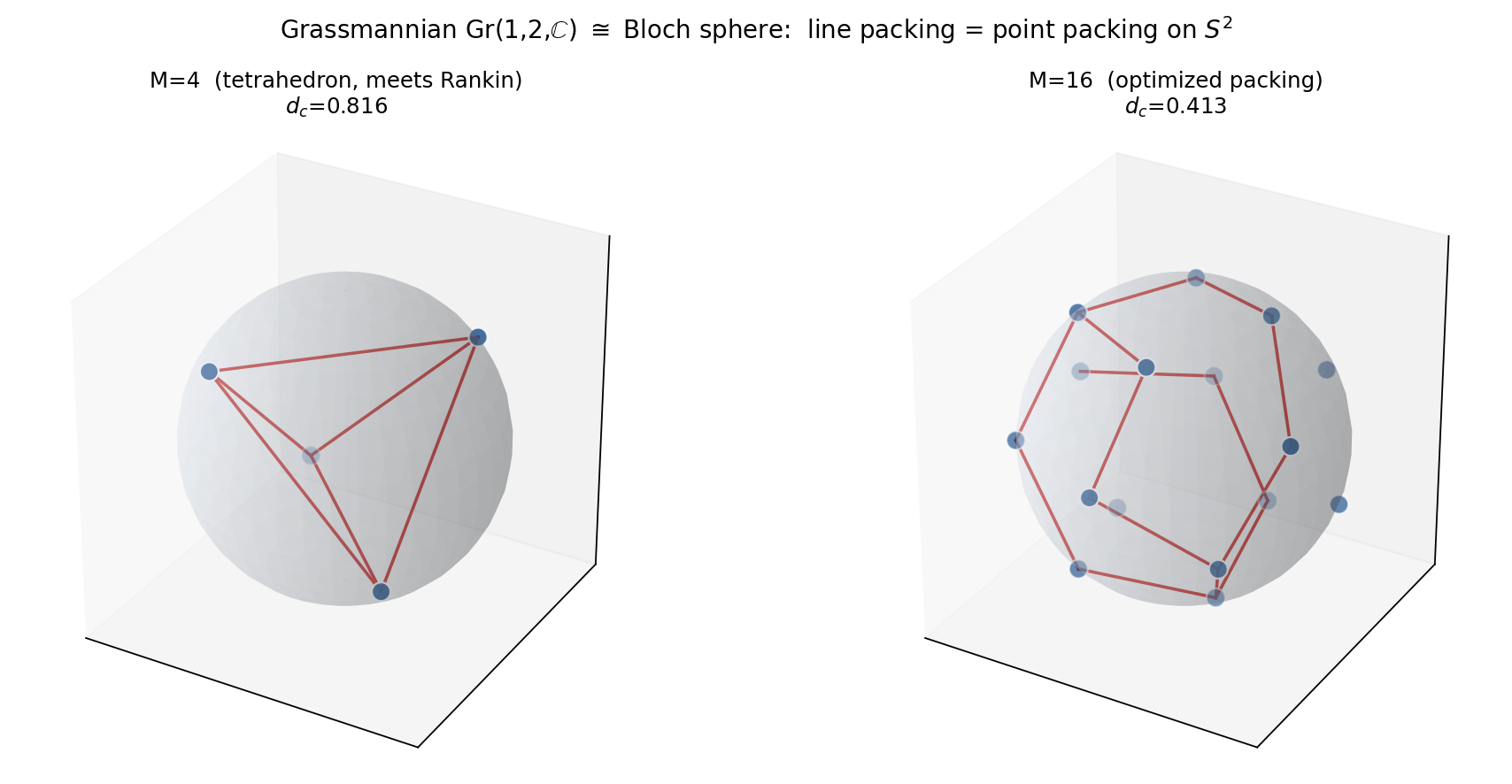}
\caption{For $T=2$, $\Gr(1,2,\C)\cong$ the Bloch sphere, so a line packing is a
sphere packing under chordal distance; the optimal $M{=}4$ packing is the
regular tetrahedron (a SIC-POVM in $\C^2$). The constellations used in this
paper live in $\Gr(1,4,\C)$ ($M{=}64$); the $T{=}2$ case is shown only for
visualization.}
\label{fig:sphere}
\end{figure}

We use $T=4$ and $M=64$ ($R=1.5$ bits/c.u.), with the constellation obtained
by minimum-coherence (max-$d_{c,\min}$) gradient descent on the manifold---
equivalently, seeking an approximate equiangular tight frame \cite{strohmer}---
giving $d_{c,\min}=0.69$; Fig.~\ref{fig:sphere} illustrates the low-dimensional
($T=2$) analogue. The block length $T=4$ balances packing freedom against rate;
the methods below are independent of $T,M$.

\paragraph*{Pulse shaping and timing}
Each block is a stream of $T$ symbols, root-raised-cosine (RRC) shaped with
roll-off $\beta$ at $L$ samples/symbol. Writing $g(\cdot)$ for the
RRC--matched-filter (raised-cosine) cascade, the symbol-rate matched-filter
output sampled with a normalized timing error $\epsilon$ (in symbols) is
\begin{equation}
y[k]=h\sum_{j} s[j]\,g\!\big(k-j-\epsilon\big)+n[k],
\label{eq:isi}
\end{equation}
which is intersymbol-interference-free only at $\epsilon=0$ ($g$ is Nyquist).

\section{PAPR-Constrained Grassmannian Constellations}\label{sec:papr}
Standard packings maximize $d_{c,\min}$ under an \emph{average}-power
constraint and place no restriction on the modulus of the individual entries
$x_t$. After pulse shaping, unequal $|x_t|$ produces large envelope
excursions; these are clipped by the Mach--Zehnder modulator's nonlinear
transfer characteristic and converted by the fiber Kerr effect into nonlinear
distortion---precisely the penalty that low-PAPR optical formats are designed
to avoid.

\begin{figure}[t]
\centering
\includegraphics[width=\columnwidth]{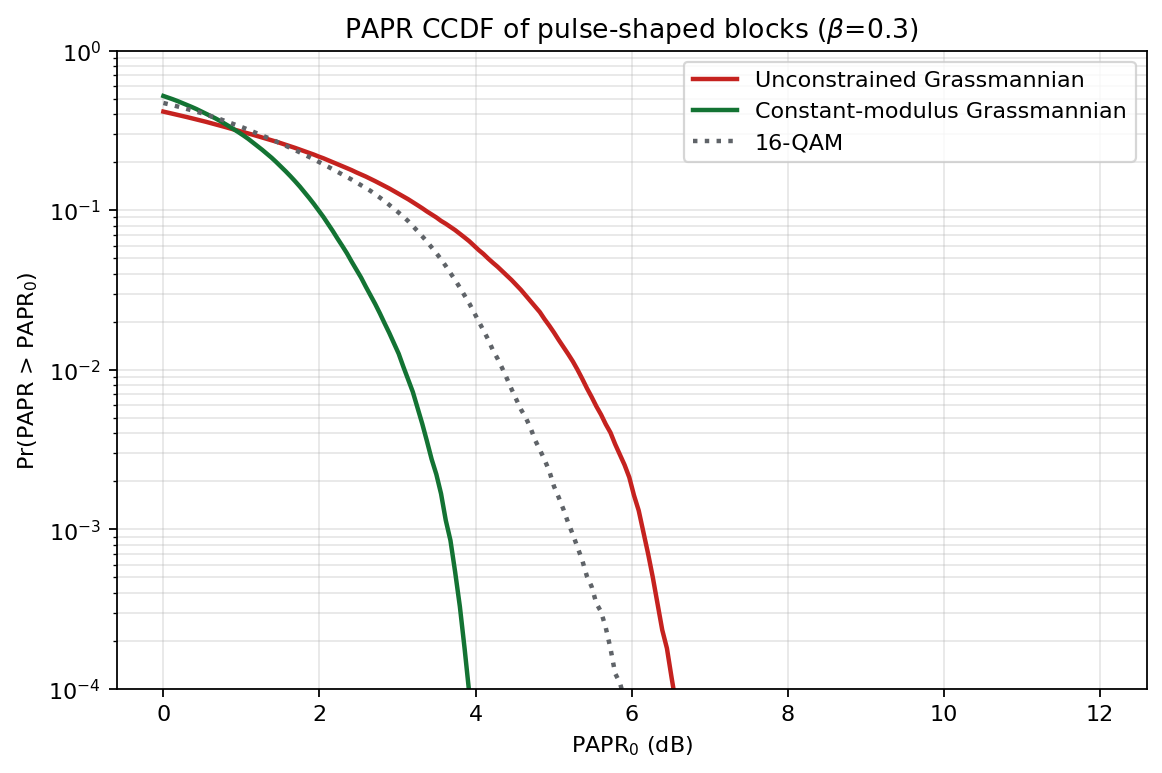}
\caption{PAPR CCDF of pulse-shaped blocks ($\beta=0.3$). The constant-modulus
Grassmannian constellation reduces the $0.1\%$ PAPR by $\sim\!2.5$~dB relative
to the unconstrained packing and by $1.6$~dB relative to $16$-QAM.}
\label{fig:papr}
\end{figure}

We therefore restrict the constellation to \emph{constant modulus},
$|x_t|=1/\sqrt{T}\ \forall t$, optimizing only the entry phases to minimize
coherence. Constant modulus equalizes the per-slot power but does not produce a
constant-envelope waveform: inter-symbol transitions through the RRC filter
still create envelope excursions, so a residual PAPR remains, which we measure
directly on the pulse-shaped blocks. Fig.~\ref{fig:papr} shows the PAPR
complementary CDF of the pulse-shaped blocks ($\beta=0.3$). The
constant-modulus Grassmannian constellation attains a $0.1\%$ PAPR of
$3.6$~dB, versus $6.1$~dB for the unconstrained packing and $5.2$~dB for
$16$-QAM---i.e., $\sim\!2.5$~dB below the unconstrained Grassmannian design and
$1.6$~dB below $16$-QAM. The price is a reduction of $d_{c,\min}$ from $0.69$
to $0.56$, about $1.8$~dB in the high-SNR coding gain. For a nonlinearity-limited
link this is expected to be a favorable trade---the reduced PAPR lowers the
required modulator back-off, which should improve nonlinear tolerance---though
quantifying the net benefit requires the split-step fiber and modulator model
deferred to future work; the constant-modulus structure additionally simplifies
driver and modulator design.

\section{Phase-Blind Subspace Timing Recovery}\label{sec:timing}
\subsection{Why a subspace-matched detector is needed}
\emph{Decision-directed} timing-error detectors \cite{mengali,rice} form the
error from the \emph{complex}, carrier-corrected sample value, which
presupposes a recovered carrier phase; in a non-coherent receiver the phase of
$\bh$ is deliberately never estimated, so they are inapplicable. Classical
\emph{non-data-aided} square-law detectors (the squaring/Oerder--Meyr
\cite{oerder} and Gardner \cite{gardner} families) are already phase-blind and
could be applied to the raw matched-filter stream, but they extract a
symbol-rate timing tone from the total energy $|y|^2$ and are agnostic to the
constellation; they do not exploit the subspace structure that defines USTM
detection. The key observation we use is that the GLRT decision statistic
\eqref{eq:glrt} is itself invariant to the channel phase: for any $\theta$,
$|\bx_m^{\mathsf H}(e^{j\theta}\by_n)|^2=|\bx_m^{\mathsf H}\by_n|^2$. The same
projection energy that drives detection can therefore drive timing recovery
without a phase reference, giving a detector \emph{matched to the subspace}
rather than to a symbol decision.

\subsection{Proposed early--late subspace TED}
Let $\by_n(\epsilon)$ denote the block of matched-filter samples taken with
timing offset $\epsilon$, and let $\hat{\bx}$ be the detected codeword from the
current (on-time) samples via \eqref{eq:glrt}. Define the projection energy
$J(\epsilon)=\sum_{n}|\hat{\bx}^{\mathsf H}\by_n(\epsilon)|^2$. Since $g$ is
Nyquist [cf.~\eqref{eq:isi}], $J$ is maximized at the correct sampling instant;
misalignment injects ISI that erodes the alignment with the detected subspace.
We form a phase-blind early--late error signal
\begin{equation}
S(\epsilon)=\sum_{n=1}^{N}
\Big[\,\bigl|\hat{\bx}^{\mathsf H}\by_n(\epsilon-\Delta)\bigr|^2
       -\bigl|\hat{\bx}^{\mathsf H}\by_n(\epsilon+\Delta)\bigr|^2\Big],
\label{eq:ted}
\end{equation}
with early--late spacing $\Delta$ (here $\Delta=0.5$). Because each term is a
squared magnitude, $S(\epsilon)$ inherits the phase invariance of
\eqref{eq:glrt}; under correct decisions the unknown $|h_n|$ enters only as a
common positive scale and does not move the zero. As
$S(\epsilon)\!\propto\!-J'(\epsilon)$ near the peak, the mean of
\eqref{eq:ted} is an S-curve with a stable zero at $\epsilon=0$ and positive
slope, suitable for a second-order tracking loop (loop filter + interpolator);
acquisition is handled by the decision-free metric of Sec.~\ref{sec:timing}-C.

\begin{figure}[t]
\centering
\includegraphics[width=\columnwidth]{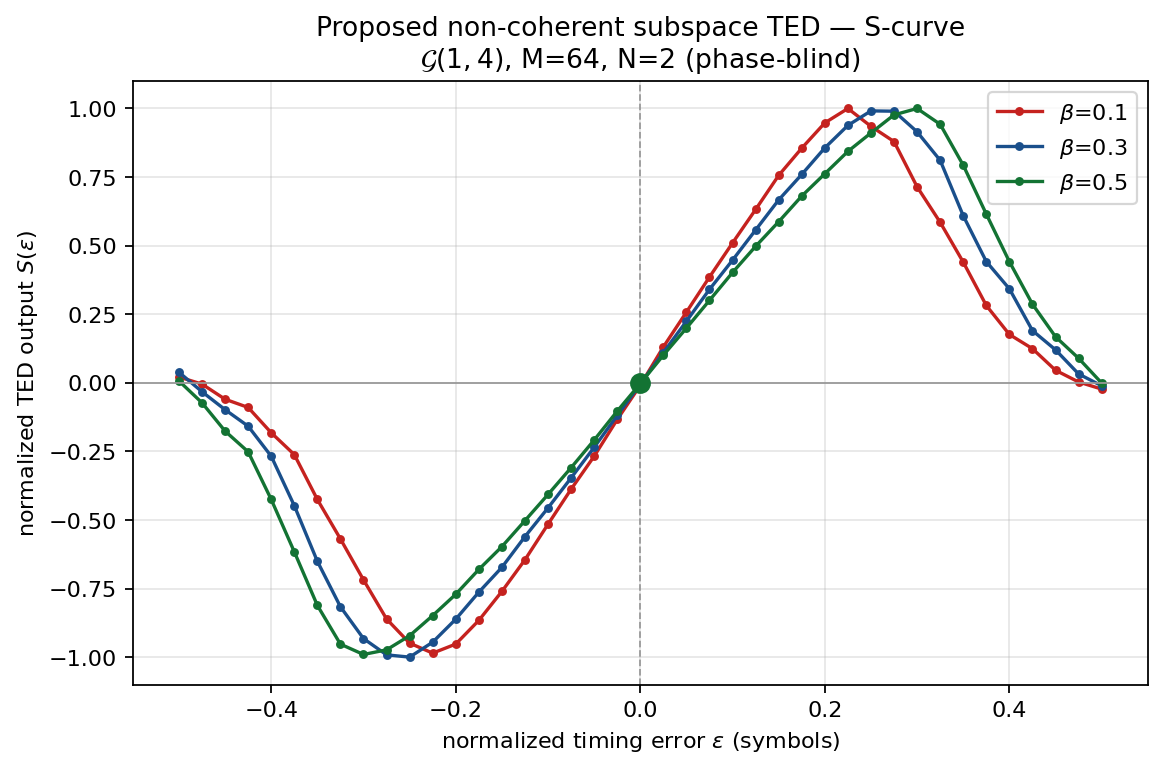}
\caption{Mean S-curve of the proposed phase-blind subspace TED \eqref{eq:ted}
for $\Gr(1,4,\C)$, $M=64$, $N=2$, shown as the noiseless mean with
unit-magnitude branch gains ($|h_n|{=}1$), normalized to unit peak. Stable zero
crossing with positive slope for all roll-offs; the normalized slope steepens
as $\beta$ decreases.}
\label{fig:scurve}
\end{figure}

Fig.~\ref{fig:scurve} plots the measured S-curve for $\beta\in\{0.1,0.3,0.5\}$.
The lock point is stable and the linear region spans roughly $\pm0.2$ symbols;
as expected, lower roll-off yields a steeper normalized slope through the lock
point and a correspondingly narrower linear range---directly relevant to the
low-roll-off regime used to maximize spectral containment.

\subsection{Feedforward acquisition}
For acquisition we use a non-data-aided variant \cite{oerder} that needs no
decisions: the
sampling phase $\phi$ is chosen to maximize the summed peak projection energy
over a block batch,
\begin{equation}
\hat{\phi}=\argmax_{\phi}\ \sum_{\text{blocks}}\ \sum_{n=1}^{N}
\max_{m}\bigl|\bx_m^{\mathsf H}\by_n(\phi)\bigr|^2 .
\label{eq:ff}
\end{equation}
Like \eqref{eq:ted}, \eqref{eq:ff} is phase-blind and exploits the
constellation structure through the inner maximization.

\section{Results and Discussion}\label{sec:results}
\begin{figure}[t]
\centering
\includegraphics[width=\columnwidth]{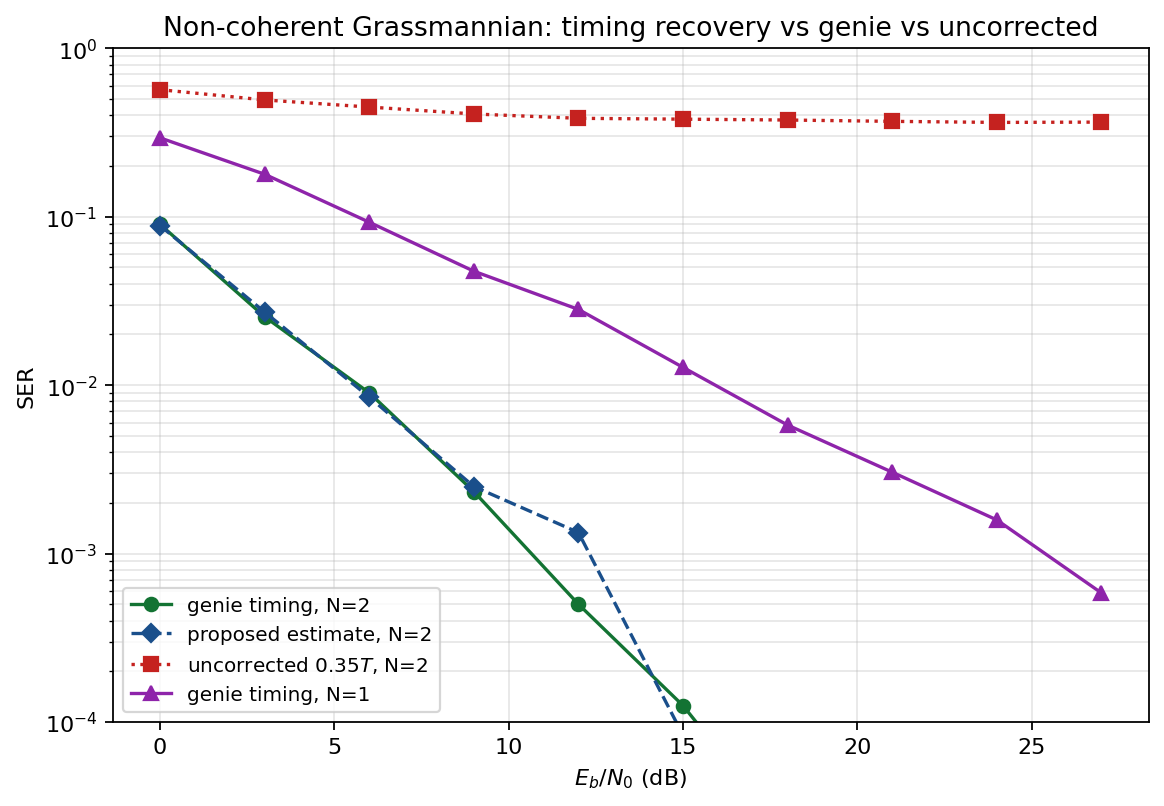}
\caption{SER over flat Rayleigh block fading ($\beta=0.3$, $T=4$, $M=64$). The
proposed phase-blind estimator matches genie timing and retains order-$N$
diversity; an uncorrected $0.35$-symbol offset floors near $0.4$.}
\label{fig:ber}
\end{figure}

We evaluate \eqref{eq:channel} with flat Rayleigh block fading, RRC
$\beta=0.3$, and the $T=4,M=64$ constellation---the unconstrained packing of
Sec.~\ref{sec:model}; the constant-modulus variant of Sec.~\ref{sec:papr}
carries the additional $\sim\!1.8$~dB $d_{c,\min}$ penalty quantified there and
is not separately simulated. Each SER point averages at least
$6\times10^3$ blocks, with the matched-filter output modeled at symbol rate via
the closed-form raised-cosine cascade over a $\pm10$-symbol ISI span.
The reported timing curves use the feedforward estimator \eqref{eq:ff}; the
S-curve of \eqref{eq:ted} is the design basis for a closed-loop tracker, left
to future work. Fig.~\ref{fig:ber} reports
symbol-error rate versus $E_b/N_0$ for: (i) genie timing; (ii) the proposed
feedforward estimator \eqref{eq:ff}; and (iii) an uncorrected $0.35$-symbol
offset, for $N=1,2$.

Three observations. (1) \emph{Timing recovery is essential}: an uncorrected
$0.35$-symbol offset destroys the subspace structure and the error rate floors
near $0.4$, independent of SNR. (2) \emph{The proposed estimator is effectively
genie}: its curve overlays genie timing within a fraction of a dB across the
whole range. (3) \emph{Diversity is preserved}: $N=2$ exhibits the expected
order-$2$ slope---at $E_b/N_0=8$~dB its SER is more than an order of magnitude
below $N=1$, the gap widening with SNR as the diversity order increases---
confirming that timing recovery does not forfeit the receive-diversity benefit
that motivates the non-coherent approach. As noted in Sec.~\ref{sec:model},
this order-$N$ diversity is an i.i.d.-Rayleigh idealization; correlated
near-unitary polarization/mode coupling would reduce it.

\paragraph*{Relation to coherent QAM}
Against conventional coherent QAM with full carrier and polarization recovery,
the proposed scheme trades spectral efficiency for robustness: at $T=4,M=64$ it
carries $1.5$~bits/c.u. and pays the usual non-coherent penalty, but it is
immune to a constant per-block carrier phase, dispenses with carrier and
polarization recovery within the block, and harvests diversity without pilots,
while its constant-modulus form eases the modulator and nonlinear budget. We do
not simulate a QAM error-rate baseline here; this comparison is qualitative
positioning, and a like-for-like comparison (matched net rate, with
blind-phase-search carrier recovery under a Wiener phase-noise process) is left
to future work. It is therefore most attractive when the link is phase-noise-
or fading-limited rather than bandwidth-limited, e.g., short-reach and access
links with wide-linewidth lasers, or mode-coupled SDM.

\section{Conclusion and Future Work}
We presented a non-coherent Grassmannian modulation for optical digital
communications, obtained by (i) constraining the constellation to constant
modulus for nonlinearity tolerance and (ii) supplying the phase-blind
symbol-timing recovery that the Grassmannian literature has lacked. Simulation
shows a clean TED S-curve, near-genie error rate after timing estimation, full
diversity, and a $1.6$--$2.5$~dB PAPR advantage.

Several items remain before submission to a measurement-grade venue and are
the subject of ongoing work: (a) replacing the block-fading abstraction with
split-step fiber propagation including the Kerr nonlinearity, a Mach--Zehnder
modulator model, and a Wiener laser-phase-noise process---which is also what is
needed to quantify the residual phase-drift floor and the nonlinear-tolerance
trade discussed above; (b) a properly tuned coherent-QAM baseline with standard
DSP (CD compensation, polarization demultiplexing, and blind-phase-search
carrier recovery \cite{pfau}) for a like-for-like spectral-efficiency
comparison, alongside a classical non-data-aided (Oerder--Meyr) TED baseline to
isolate the gain from the subspace-matched construction; (c) a full closed-loop
timing tracker (loop filter + NCO + interpolator) under time-varying offset, of
which the S-curve here is the design basis; (d) a correlated, near-unitary
polarization/SDM coupling channel in place of i.i.d.\ Rayleigh; and (e)
experimental validation on a coherent or SDM testbed. We also note that the
present link-level results use a symbol-rate AWGN abstraction after matched
filtering; the qualitative ordering (estimated $\approx$ genie $\gg$
uncorrected, and order-$N$ diversity) is expected to persist under the full
continuous-time model.

\section*{Reproducibility}
The simulation code reproducing Figs.~\ref{fig:papr}--\ref{fig:ber} is
available from the author on request.

\end{document}